\documentclass{IEEEcsmag}

\usepackage[colorlinks,urlcolor=blue,linkcolor=blue,citecolor=blue]{hyperref}
\expandafter\def\expandafter\UrlBreaks\expandafter{\UrlBreaks\do\/\do\*\do\-\do\~\do\'\do\"\do\-}
\usepackage{upmath,color}
\usepackage{booktabs}
\usepackage{framed}

\jvol{XX}
\jnum{XX}
\paper{8}
\jmonth{Month}
\jname{Publication Name}
\jtitle{Publication Title}
\pubyear{2023}

\begin{document}

\sptitle{Feature: Documentation Practices of Machine Learning Resources}

\title{The State of Documentation Practices of Third-party Machine Learning Models and Datasets}

\author{Ernesto Lang Oreamuno}
\affil{Queen's University, Canada}

\author{Rohan Faiyaz Khan}
\affil{Queen's University, Canada}

\author{Abdul Ali Bangash}
\affil{Queen's University, Canada}

\author{Catherine Stinson}
\affil{Queen's University, Canada}

\author{Bram Adams}
\affil{Queen's University, Canada}

\markboth{FEATURE}{FEATURE}

\begin{abstract}
\looseness-1 Model stores offer third-party ML models and datasets for easy project integration, minimizing coding efforts. One might hope to find detailed specifications of these models and datasets in the documentation, leveraging documentation standards such as model and dataset cards. 
In this study, we use statistical analysis and hybrid card sorting to assess the state of the practice of documenting model cards and dataset cards in one of the largest model stores in use today--Hugging Face (HF).
Our findings show that only 21,902 models (39.62\%) and 1,925 datasets (28.48\%) have documentation. Furthermore, we observe inconsistency in ethics and transparency-related documentation for ML models and datasets.

\end{abstract}

\maketitle
\FrameSep4pt
\label{sec:introduction}
\chapteri{M}achine learning (ML) model stores like HF (valued at \$80 million USD)\footnote{https://www.forbes.com/sites/kenrickcai/2022/05/09/the-2-billion-emoji-hugging-face-wants-to-be-launchpad-for-a-machine-learning-revolution/}) offer ML models and datasets for reuse and deployment. These stores, similar to smartphone app stores and third-party libraries, furnish instructions and documentation for integrating and using ML models or reusing datasets.

Inadequate documentation for ML models and datasets, similar to third-party libraries, can have significant consequences. The absence of clear guidelines and usage instructions hinders reusability. The lack of comprehensive documentation obscures ethical concerns, biases, and limitations, increasing the risk of ethical violations and bias.
For instance, Rajbahadur~\textit{et al.}~\cite{gopi23} suggest potential risks of license violations in the datasets of ML models.
Moreover, in a recent study, Jiang et al.~\cite{jiang2023empirical} faced difficulty in reasoning about model reusability because 80\% of the HF models were insufficiently documented. To address these concerns, ML stores like HF introduced Model Cards and Dataset Cards for documentation.

A \textit{Model card}, adhering to the standard template~\cite{mitchell_model_2019}, governs model documentation, encompassing general model descriptions, intended use, performance evaluation metrics, etc.
Similarly, a \textit{Dataset card}, adhering to the standard template~\cite{mcmillan-major_reusable_2021}, regulates dataset documentation, covering general dataset descriptions, structure (including data fields and splits), and considerations for data use (like biases and known limitations).

To date, the availability and quality of the documentation of ML models and their datasets have not been studied extensively.
Bhat et al.~\cite{bhat2023aspirations} investigated the state of the documentation of models. However, their analysis was limited to 50 model cards on the HF models store. 
No work thus far has studied the documentation quality of datasets in online stores.

In this work, we qualitatively study 378 model cards and 321 dataset cards available on HF. Specifically, we investigate the prevalence of model and dataset cards, what kind of information the model/data engineers use in the HF store to document their models and datasets, and how much these cards comply with the current ML documentation standards~\cite{mitchell_model_2019,mcmillan-major_reusable_2021}.

Our findings carry implications for researchers, model deployers, and tool builders. Given the escalating dependence on third-party models and datasets within software engineering, the significance of precise documentation continues to amplify. Our dataset is publicly available for researchers\footnote{https://zenodo.org/records/10256770}.



\section{DOCUMENTATION PRACTICES}
\label{sec:backg:subsec:se_docs}

Nahar~et~al. hinted at the sparse documentation of datasets, but without empirical evidence~\cite{nahar_collaboration_2021}. Fischer~\textit{et al.} highlighted the challenges of documenting models and datasets~\cite{fischer_system_2021}. 
Bandy~et~al.~\cite{bandy_documentation_debt_2020} proposed datasheets for datasets and found problems ranging from copyright issues, duplication, and unmarked biases in the documentation of the BookCorpus dataset.

Bhat~\textit{et al.}~\cite{bhat2023aspirations} performed a qualitative investigation on 50 HF Model Cards.
We contrast our analysis from them with 378 HF Model Cards and 321 HD Dataset Cards, which have previously not been studied in the literature.
Similar to our conclusion, Bhat et al.~\cite{bhat2023aspirations} found ethical considerations least frequently addressed. Moreover, they found contact information, model license, and performance visuals to be sparsely documented, whereas, in addition to their results, we found metrics, citation details, caveats, and recommendations to be sparsely documented in the Model Cards.
\section{METHODOLOGY}
\label{sec:methodology}

\subsubsection{Model Card and Dataset Card Scraping}
Initially, we collected all model cards for the 55,280 models and all dataset cards for the 6,758 datasets present in HF at the time of this research (July 4, 2022). 
For the remainder of the study, we filtered out cards that were empty, leaving 21,902 model cards and 1,925 dataset cards.

\subsubsection{Hybrid Card Sorting}

Two authors conducted hybrid card sorting on a random sample of 378 model cards with 95\% confidence (5\% confidence interval) to understand their content. This process involved categorizing card sections according to model and dataset card standards~\cite{mitchell_model_2019,mcmillan-major_reusable_2021} to determine if they contained the expected information. 
For example, a section titled \textit{Model Description} would be matched with the \textit{Model Details} section of the standard. Sections that did not precisely match their titles or appropriate categories in the standards were labeled as having mismatched content.
Our Cohen's Kappa inter-rater agreement reached 79\% during hybrid card sorting, with discrepancies resolved through consensus.

\subsubsection{Deductive Coding}
We also conducted deductive coding~\cite{beyer_manual_2014} based on the sections and subsections outlined in the dataset card standards~\cite{mcmillan-major_reusable_2021}. For instance, in a card-subsection like \textit{Dataset Summary}, we ensured that (1) its content matched the intended documentation and (2) the title and content were appropriately aligned. This assessment involved two authors analyzing a random sample of 321 datasets chosen with 95\% confidence level (and 5\% confidence interval). The authors achieved a Cohen's Kappa inter-rater agreement rate of 91\%. In instances of disagreement, they collaborated to discuss and reach a consensus.

\section{RESULTS: MODEL CARDS}
\label{sec:rq1-results}

\textbf{60.38\% of 55,280 models in HF do not have a model card.} 
This phenomenon aligns with previous research findings that identified the absence of documentation as a significant concern~\cite{aghajani_software_doc_2019}. While software engineering documentation also tends to lack installation, deployment and release instructions~\cite{aghajani_software_2020}, their absence for ML models (which are more abstract) might even be a larger concern. 

\begin{figure}[ht]
    \centering
    \includegraphics[width=\linewidth]{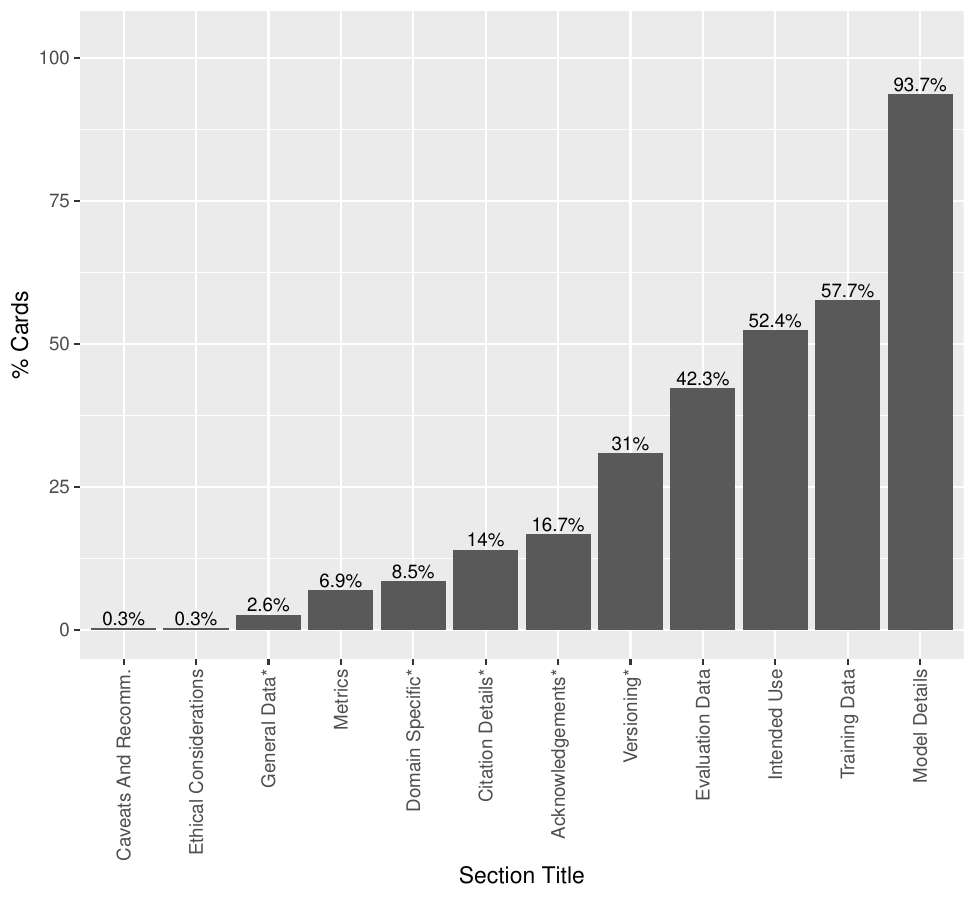}
    \caption{Model card section title prevalence across the 378 sampled model cards. A total of 2,053 sections were analyzed, out of which 19 remain un-categorized.}
    \label{fig:rq1-cats}
\end{figure}

\textbf{At the model card section-title level, more than half of the HF model cards document the \textit{Model Details}, \textit{Training Data}, and \textit{Intended Use} sections}. Figure \ref{fig:rq1-cats} displays a bar chart with the percentage of occurrences of each model card section title throughout our sample of 378 model cards. This figure shows 12 section titles, out of which 5 are not part of the official model card standard~\cite{mitchell_model_2019}, but were identified during our manual analysis of the section titles. In the figure, these additional sections include a star* along with their names.

Model providers at HF are especially focused on providing information about \textit{Model Details} (93.7\%), \textit{Training Data} (57.7\%), and \textit{Intended Use} (52.4\%). Surprisingly, documentation of categories such as \textit{Metrics} appear less common, with only 6.9\% occurrence. Particularly concerning is the lack of documentation concerning \textit{Ethical Considerations} and \textit{Caveats and Recommendations} (0.3\% for both).

\begin{figure}[ht]
    \centering
    \includegraphics[width=\linewidth]{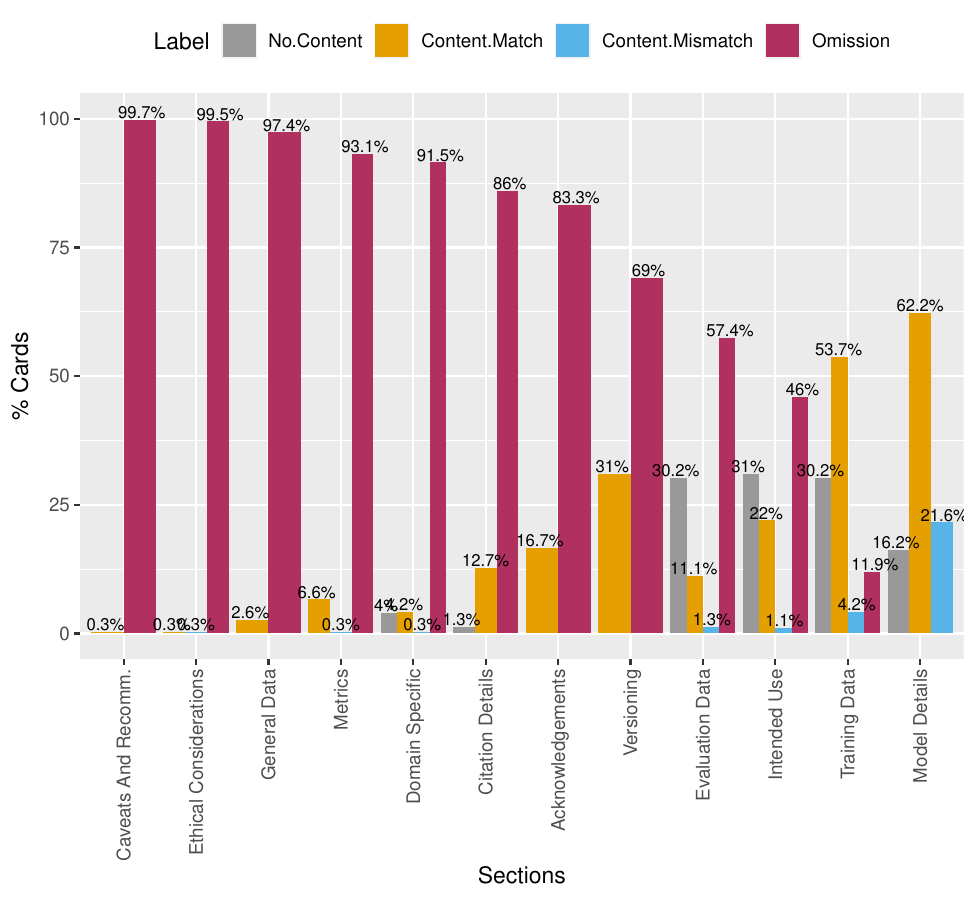}
    \caption{Content (Mis-)Match prevalence in the section-categories across 378 Model Cards. Omission means a sub-section was not present in the documentation.}
    \label{fig:rq1-manual}
\end{figure}

\textbf{\textit{Model Details} is first, while \textit{Training Data} is second in content matches. \textit{Intended Use}, \textit{Training Data} and \textit{Evaluation Data} have the most empty sections. \textit{Ethical Considerations} and \textit{Caveats And Recommendations} have almost no documentation}. 
 
Figure~\ref{fig:rq1-manual} presents our hybrid card sorting results on the content of the 378 analyzed model cards. Each bar represents a different category, with purple bars representing the \% of sections authors did not provide, grey representing the \% of sections provided but without content, blue representing the \% of sections with content that mismatch its title, and orange representing the \% of sections with content that match its title.

The most documented sections of Figure~\ref{fig:rq1-cats} show substantial inconsistencies in the quality of their content. For instance, \textit{Model Details} has a high mismatched content percentage of 21.6\%, while \textit{Intended Use} and \textit{Training Data} have a high \% of empty content with 31\% and 30.2\%, respectively. 

On average, 8.79\% of sections do not have content, 2.3\% of sections content does not match their actual specification, while 17.38\% of sections content matches the specification.
In general, the providers of models at HF are more consistent w.r.t information about \textit{Model Details} and \textit{Training Data}, while \textit{Metrics}, \textit{Ethical Considerations} and \textit{Caveats and Recommendations} have only 0.3\% to 6.6\% content that matches the specification. The lack of content match in \textit{Ethical Considerations} and \textit{Caveats and Recommendations} is worrying.

Consider the case of MikePence~\footnote{https://huggingface.co/huggingtweets/mike\_pence/}, a model that generates tweets based on the historical tweets of the former Vice President of the United States. By leaving out \textit{Ethical Considerations} and \textit{Caveats and Recommendations}, people using this model in a production context might face legal consequences.
Even large companies can fail to attend to the legalities of the code and data they acquire, as demonstrated by Google DeepMind, who faced fines for deploying a model that used NHS dataset~\cite{paleyes_challenges_2021}. 

\begin{framed}
\noindent
60.38\% of model cards are not documented. On average, 8.79\% of the sections do not have content, and 2.3\% of the content does not match the model card standards. Model cards are focused mostly on \textit{Model Details}, \textit{Training Data}, and \textit{Intended Use}, while \textit{Ethical Considerations} and \textit{Caveats and Recommendations} are rarely documented.
\end{framed}

\section{RESULTS: DATASET CARDS}
\label{sec:rq2-results}

\textbf{71.52\% of the 6,758 datasets do not have a dataset card.}

\begin{figure}[ht]
    \centering
    \includegraphics[width=\linewidth]{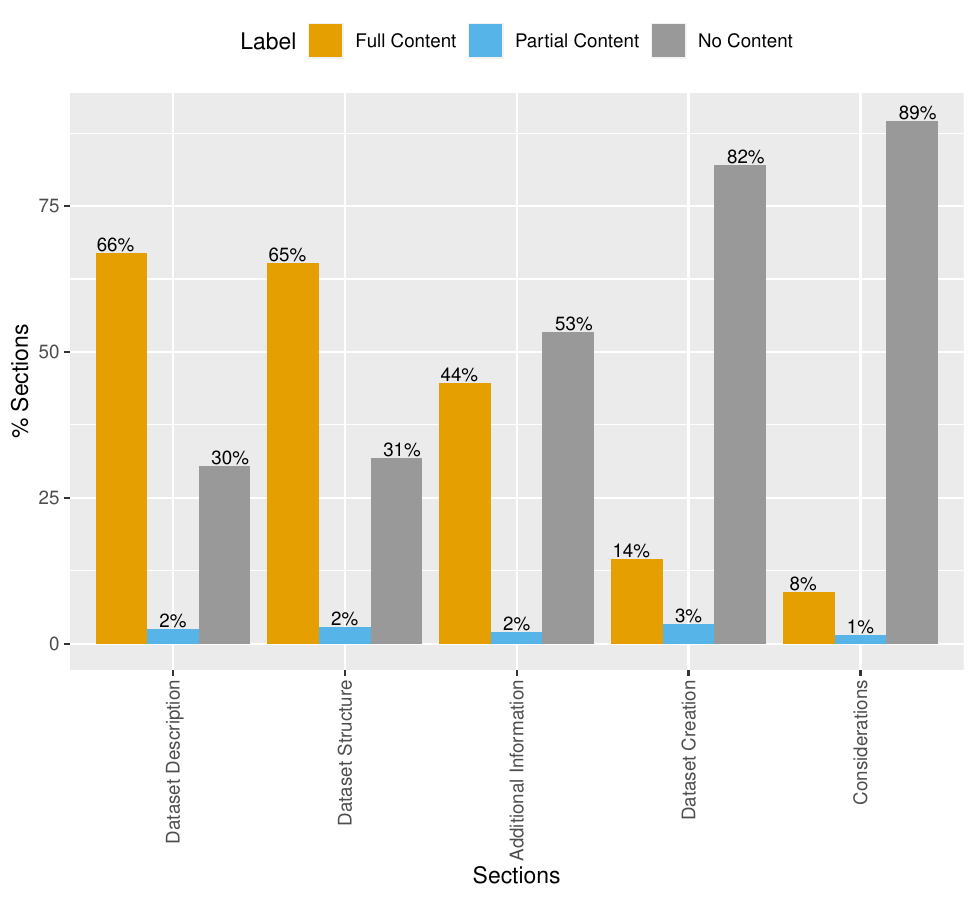}
    \caption{Completeness of the content of Dataset Card sections. Percentage values rounded-up to IEC 60559 standard for presentation.}
    \label{fig:rq2-dataset-sections}
\end{figure}

\textbf{At the dataset card section-title level, the standards used to document dataset cards are followed inconsistently}. Figure \ref{fig:rq2-dataset-sections} provides a bar chart with the section title-level percentages of fully documented content, partially documented content and lack of content for all 9,622 sections of the 321 sampled dataset cards. 

Even though HF enforces the use of the dataset card standard~\cite{mcmillan-major_reusable_2021} (i.e., there are no top-level sections omitted here), in contrast to model cards, the \textit{Dataset Description} and the \textit{Dataset Structure} sections are more documented than the other sections, presenting 68\% and 67\% across the dataset cards (when adding up full and partial content), compared to 46\% for \textit{Additional Information}, 17\% for \textit{Dataset Creation} and 9\% for \textit{Considerations}.

A positive observation is that all the sections have a low percentage of partial content instances (1\% to 3\%) compared to full content (8\% to 66\%). \textit{Dataset Description} and \textit{Dataset Structure} have more full content (66\% and 65\%) than no content (30\% and 31\%), whereas the other three sections, follow an opposite trend where ``No Content'' surpasses the total available content. This trend is worrying, since \textit{Dataset Creation} and \textit{Considerations For Using The Data} are sections that are key to understanding potential copyright violations~\cite{gopi23}. 

\begin{figure}[ht]
    \centering
    \includegraphics[width=\linewidth]{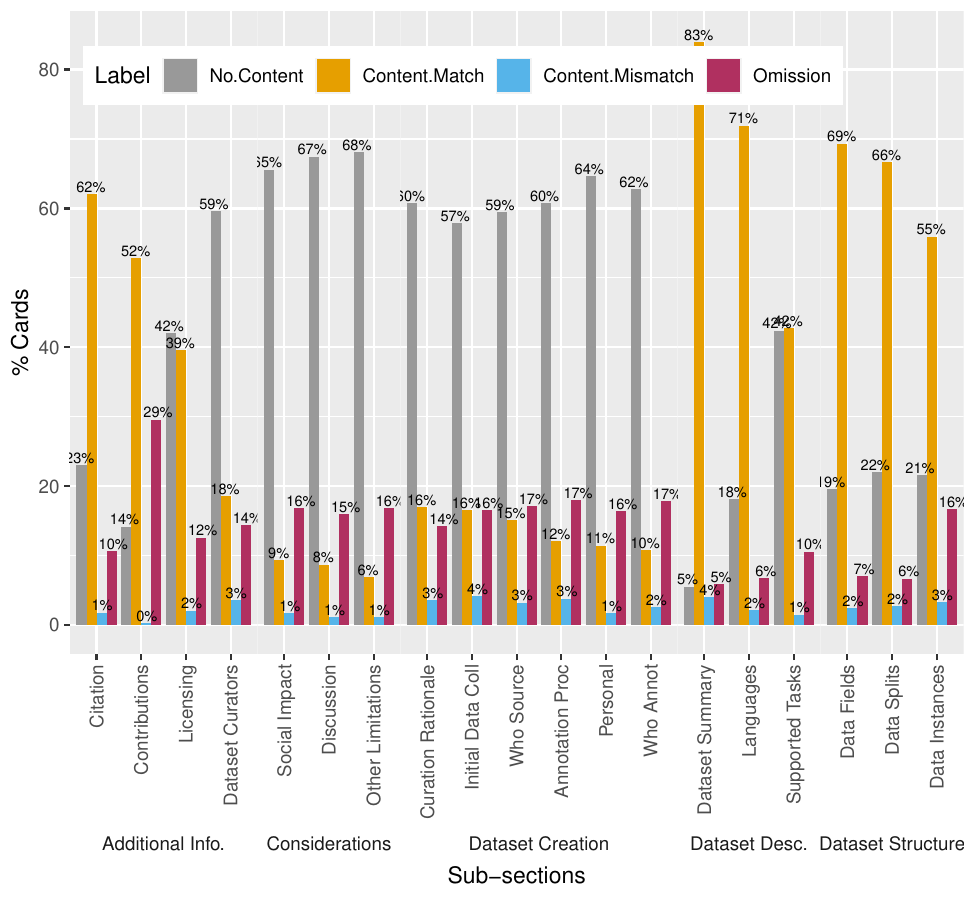}
    \caption{Content (Mis-)Match prevalence in the sub-sections of 321 Dataset Cards. Percentage values rounded-up to IEC 60559 standard for presentation.}
    \label{fig:rq2-manual}
\end{figure}

\textbf{At the dataset card subsection level, our deductive coding of the content of the 321 dataset cards shows that more than half (10/19) of the subsections are empty in at least 57\% of the cards, while less than half (7/19) have at least 52\% content match.} Figure~\ref{fig:rq2-manual} presents a bar chart with the results of our deductive coding of 321 cards for the content of the 19 subsections of the dataset card standard. 

\textbf{All dataset card subsections have on average 14.27\% omissions (missing sections)}.
\textit{Contributions} has the most omissions (29\%). A similar trend is followed by most of the subsections of the \textit{Dataset Creation} and \textit{Considerations} sections. 
Moreover, the \textit{Data Instances} sub-section (in \textit{Dataset Structure} section) and, to some extent, \textit{Supported Tasks} sub-section (in \textit{Dataset Description} section) have omissions in at least 10\% cards. 

To investigate if the ``dataset creators'' information is available in the datasets documentation, we observe the content of the sub-sections: \textit{Who-Annotated} and \textit{Dataset-Curators}. 
Dataset curators are usually responsible for the collection and cleaning of data, while the annotators add annotations or labels to the data to make it suitable for specific machine-learning tasks.
We find that annotators' information is available in only 12\%, while information about curators is available in 21\% data cards.
To investigate if the ``source of data'' is available in the documentation of the datasets, we observe the sub-section \textit{Who-Source} content, which is documented in 18\% of the data cards. 
We also observe the section \textit{Supported Tasks}, which explains which type of ``Machine Learning tasks'' a dataset was curated for, and the section \textit{Licensing}, which explains what are the ``legal implications'' of using a dataset. The former is documented in 43\%, while the latter is documented in 41\% of the cards only.
This comprehensive overview further emphasizes the need to improve transparency in the documentation of datasets.

To identify tools that are frequently used in the creation and annotation of the datasets, future work can manually inspect the data cards that we have collected in the following sub-categories: \textit{Curation Rationale}, \textit{Annotation Procedure}, and \textit{Initial Data Collection}.

\begin{framed}
\noindent
The majority of datasets (71.52\%) are not documented. At the subsection level, 10/19 subsections are empty in more than half of the instances, while only 7/19 have a complete match in more than half of their instances.
\end{framed}

\section{LIMITATIONS}
\label{sec:threats}

To investigate our random sampling bias, we compared the distribution of full/partial/no-match section instances across 50\% cards with the highest .vs. lowest number of downloads. Only for model cards, we found significant differences (sections ``Model Details'', ``Intended Use'', ``Training Data'', ``Evaluation Data'', and ``Domain Specific''), yet except for ``Domain Specific'' (which has few data points) the Cramer effect size indicated only moderate differences. 

Bhat~\textit{et~al.}~\cite{bhat2023aspirations} found that documentation for third-party models can also be available on other platforms, such as Github. Additionally, HF now has 350,000 models compared to the 55,280 models that were available during our large-scale qualitative study.
Therefore, our findings may not generalize over the documentation of the latest models and datasets on HF.

\label{sec:implications}

\section{CALL FOR RESEARCHERS}
\label{sec:implications:subsec:research}

Our findings indicate that the documentation practice in HF either does not meet the needs of users or is not adequately enforced. Moreover, the existing HF Model Cards lack essential categories related to model versioning and attribution.

Understanding the reasons behind these documentation shortcomings in HF Models and Datasets is essential for developing effective documentation policies. 
Researchers should investigate factors influencing ML practitioners' perceptions of documentation importance. 
Additionally, they should assess why current standards like Model Cards and Dataset Cards~\cite{mitchell_model_2019, mcmillan-major_reusable_2021}, designed to address ethical concerns, have not effectively met their primary goal. Research is needed to identify gaps in ethical considerations within documentation practices and propose improvements.

\section{CALL FOR ML-PRACTITIONERS}
\label{sec:implications:subsec:deployment}
Deploying third-party models in a business context can have legal consequences, especially due to limited ethical information. Models lacking ethical documentation may face legal scrutiny, impacting both the model and its integrated product~\cite{counter-radicalization_marcellino__2020}. 

ML practitioners should actively advocate for enhanced documentation practices within their organizations and the ML community. This can be achieved by raising awareness about the significance of thorough documentation, sharing experiences related to deployment challenges, and actively participating in discussions aimed at driving positive changes in documentation practices.

\section{CALL FOR TOOL BUILDERS}
\label{sec:implications:subsec:future}
Tool builders should focus on developing advanced documentation tools tailored to the needs of software engineers, data engineers, and data scientists in the ML domain. 
Considering the highlighted ``dataset documentation'' in this study, tool builders should create solutions that streamline the documentation process for datasets, ensuring consistency, completeness, and compliance with established standards.

Tools are available to extract documentation information automatically from source code, test code, or both~\cite{croft2023}, and to remind the ML practitioners to update their documentation~\cite{bhat2023aspirations}. Additionally, there are tools available to assess the quality of existing documentation using source code or manually curated quality metrics~\cite{henry2023}. However, the model repositories at HF differ in nature; they lack source code but instead contain machine learning models and configuration files explaining model parameters. Similarly, HF dataset repositories store datasets exclusively. However, both types of repositories reference their source code links on Github, which contain code for model training and evaluation or dataset scraping. Existing tools in the literature can be adapted to utilize the information available in the source code links on HF repositories. Furthermore, new tools can be developed to leverage information that is unique to HF dataset repositories, such as HF community discussions.

 


\def\refname{REFERENCES}

\vspace*{-8pt}

\begin{IEEEbiography}{ERNESTO L. OREAMUNO}{\,} has a Master's from Queen's University in the area of software engineering in AI or SE4AI for short. His thesis looks into how software engineering practices and projects can be combined with machine learning artifacts such as datasets and models, with the intention of exploring how these elements are coalesced together into applicable business solutions. Beyond that, he has worked in software engineering for the past 5 years, using technologies such as Docker, PostgresSQL, Nginx, RabbitMQ, Google Cloud Platform and Nginx. Currently, his time is dedicated to massive data extraction and analysis through the use of machine learning. Contact him at ernesto.lang@queensu.ca\vadjust{\vfill\pagebreak}
\end{IEEEbiography}

\begin{IEEEbiography}{ROHAN FAIYAZ KHAN}{\,} is a PhD student in Computer Science at Queen's University (Ethics and Technology Lab). Her research focuses on ethics of AI, racial biases in AI, facial recognition and medical imaging. Contact her at 21rfk@queensu.ca
\vspace*{8pt}
\end{IEEEbiography}

\begin{IEEEbiography}{ABDUL A. BANGASH}{\,} is a Postdoctoral Fellow at the SAIL lab at Queen’s University, Canada. He is privileged to be one of the six selected candidates who have been awarded the esteemed ``Vice-Principal Research (VPR) Postdoctoral Fund'' at Queen’s University. He completed his PhD in Computer Science from the University of Alberta, Canada. His research focus lies in enhancing engineering processes of software and machine-learning frameworks through techniques such as mining software repositories, search-based software engineering, static analysis, and energy optimization.
Contact him at abdulali.b@queensu.ca; http://abdulali.github.io\vspace*{8pt}
\end{IEEEbiography}

\begin{IEEEbiography}{CATHERINE STINSON} {\,} is an Assistant Professor and Queen’s National Scholar in the Philosophical Implications of Artificial Intelligence at Queen’s University. Their research interests include developing and evaluating benchmarks for generative AI, including BIG-Bench, investigating bias in AI systems, and supporting data justice projects like \url{https://trackinginjustice.ca/}. They have presented at premier AI Ethics conferences like AIES, and FAccT, and published in top philosophy of science journals like Synthese, Philosophy of Science, and AI \& Ethics.
Contact them at c.stinson@queensu.ca\vspace*{8pt}
\end{IEEEbiography}

\begin{IEEEbiography}{BRAM ADAMS} {\,} is a full professor at Queen's University. His research interests include software release engineering (pre- and post-AI) and mining software repositories. His work has received the 2021 Mining Software Repositories Foundational Contribution Award. In addition to co-organizing the RELENG International Workshop on Release Engineering from 2013 to 2015 (and the 1st/2nd IEEE Software Special Issue on Release Engineering), he co-organized the first editions of the SEMLA event on Software Engineering for Machine Learning Applications. He has been PC co-chair of SCAM 2013, SANER 2015, ICSME 2016 and MSR 2019, and ICSE 2023 software analytics area co-chair. He is a Senior IEEE Member. Contact him at bram.adams@queensu.ca
\end{IEEEbiography}


\begin{thebibliography}{1}

\bibitem{gopi23}
G.~Rajbahadur, E.~Tuck, L.~Zi, D.~Lin, B.~Chen, Z.~Jiang, and D.~German,
``Can I use this publicly available dataset to build commercial
AI software?,'' in \emph{2022 arXiv:2111.02374v5}.

\bibitem{aghajani_software_doc_2019}
E.~Aghajani, C.~Nagy, O.~L. Vega-Márquez, M.~Linares-Vásquez, L.~Moreno,
G.~Bavota, and M.~Lanza, ``Software documentation issues unveiled,'' in
\emph{2019 IEEE/ACM 41st International Conference on Software Engineering}, 2019.

\bibitem{mitchell_model_2019}
M.~Mitchell, S.~Wu, A.~Zaldivar, P.~Barnes, L.~Vasserman, B.~Hutchinson,
E.~Spitzer, I.~D. Raji, and T.~Gebru, ``Model cards for model reporting,'' in
\emph{Proceedings of the 2nd Conference on Fairness, Accountability, and
Transparency}, 2019.

\bibitem{mcmillan-major_reusable_2021}
A.~McMillan-Major, S.~Osei, J.~D. Rodriguez, P.~S. Ammanamanchi, S.~Gehrmann,
and Y.~Jernite, ``Reusable templates and guides for documenting datasets and
models for natural language processing and generation: A case study of the
HuggingFace and GEM data and model cards,'' in \emph{Proceedings of the 1st
Workshop on Natural Language Generation, Evaluation, and Metrics},
2021.

\bibitem{nahar_collaboration_2021}
N.~Nahar, S.~Zhou, G.~Lewis, and C.~K{\"a}stner, ``Collaboration challenges in
building ml-enabled systems: Communication, documentation, engineering, and
process,'' in \emph{Proceedings of the 44th International Conference on
Software Engineering}, 2022.

\bibitem{fischer_system_2021}
L.~Fischer, L.~Ehrlinger, V.~Geist, R.~Ramler, F.~Sobiezky, W.~Zellinger,
D.~Brunner, M.~Kumar, and B.~Moser, ``Ai system engineering—key challenges
and lessons learned,'' \emph{2021 Machine Learning and Knowledge Extraction}.

\bibitem{bandy_documentation_debt_2020}
J.~Bandy and N.~Vincent, ``Addressing "documentation debt" in machine learning:
A retrospective datasheet for bookcorpus,'' in \emph{Proceedings of the
Neural Information Processing Systems Track on Datasets and Benchmarks}

\bibitem{bhat2023aspirations}
A.~Bhat, A.~Coursey, G.~Hu, S.~Li, N.~Nahar, S.~Zhou, C.~Kastner, and 
J.~Guo, ``Aspirations and Practice of ML Model Documentation: Moving the Needle with Nudging and Traceability,''
in \emph{2023 Proceedings of the CHI Conference on Human Factors in Computing Systems}.

\bibitem{beyer_manual_2014}
S.~Beyer and M.~Pinzger, ``A manual categorization of android app development
issues on stack overflow,'' in \emph{2014 IEEE International Conference on
Software Maintenance and Evolution}.

\bibitem{aghajani_software_2020}
E.~Aghajani, C.~Nagy, M.~Linares-Vásquez, L.~Moreno, G.~Bavota, M.~Lanza, and
D.~C. Shepherd, ``Software documentation: The practitioners' perspective,''
in \emph{2020 IEEE/ACM 42nd International Conference on Software Engineering}.

\bibitem{croft2023}
R.~Croft, M.~Babar, and M.~Kholoosi, 
``Data Quality for Software Vulnerability Datasets''
in \emph{2023 IEEE/ACM 45th International Conference on Software Engineering}.

\bibitem{henry2023}
H.~Tang, and S.~Nadi,
``Evaluating Software Documentation Quality,''
in \emph{2023 IEEE/ACM 20th International Conference on Mining Software Repositories }.

\bibitem{paleyes_challenges_2021}
A.~Paleyes, R.-G. Urma, and N.~Lawrence, ``Challenges in deploying machine
learning: A survey of case studies,'' in \emph{Proceedings of the Working on
ML-Retrospectives, Surveys \& Meta-Analyses (ML-RSA), co-located with the
34th Conference on Neural Information Processing Systems}.

\bibitem{counter-radicalization_marcellino__2020}
W.~Marcellino, M.~Magnuson, A.~Stickells, B.~Boudreaux, T.~Helmus, E.~Geist,
and Z.~Winkelman, \emph{Counter-Radicalization Bot Research: Using Social
Bots to Fight Violent Extremism}.\hskip 1em plus 0.5em minus 0.4em\relax RAND
Corporation, 2020.

\bibitem{jiang2023empirical}
W.~Jiang, N.~Synovic, M.~Hyatt, T.~Schorlemmer, R.~Sethi, Y.~Lu and G.~Thiruvathukal, J.~Davis, 
``An empirical study of pre-trained model reuse in the hugging face deep learning model registry,''
in \emph{2023 IEEE/ACM 45th International Conference on Software Engineering}.

\end{thebibliography}
\end{document}